\documentclass[preprint2]{aastex}

\shorttitle{Orbital Dynamics: GJ 832 System}
\shortauthors{Satyal et. al.}

\begin{document}


\title{Dynamics of a Probable Earth-mass Planet in GJ 832 System}

\author{S. Satyal$^1$\altaffilmark{*}, J. Griffith$^1$, Z. E. Musielak$^{1,2}$}
\affil{$^1$The Department of Physics, University of Texas at Arlington, Arlington, TX 76019.}
\affil{$^2$Kiepenheuer-Institut f\"ur Sonnenphysik, Sch\"oneckstr. 6, 79104 Freiburg, Germany.}

\altaffiltext{*}{corresponding author: ssatyal@uta.edu}

\begin{abstract}
Stability of planetary orbits around the GJ 832 star system, which contains inner (GJ 832c) and outer (GJ 832b) planets, is investigated numerically and a detailed phase-space analysis is performed.  A special emphasis is given to the existence of stable orbits for a planet less than 15M$_\oplus$ which is injected between the inner and outer planets. Thus, numerical simulations are performed for three and four bodies in elliptical orbits (or circular for special cases) by using a large number of initial conditions that cover the selected phase-spaces of the planet's orbital parameters. The results presented in the phase-space maps for GJ 832c indicate the least deviation of eccentricity from its nominal value, which is then used to determine its inclination regime relative to the star-outer planet plane. Also, the injected planet displays stable orbital configurations for at least one billion years. Then, the radial velocity curves based on the signature from the Keplerian motion are generated for the injected planets with masses 1M$_\oplus$ to 15M$_\oplus$ in order to estimate their semimajor axes and mass-limit. The synthetic RV signal suggests that an additional planet of mass $\le$ 15M$_\oplus$ with dynamically stable configuration may be residing between 0.25 - 2.0 AU from the star. We have provided an estimated number of RV observations for the additional planet that is required for further observational verification.
\\

\\
\end{abstract}

\section{Introduction}

Recent discovery of exoplanets has shown that the multi-planetary systems in compact orbits seem common in the Milky Way Galaxy. For example, Kepler 186 \citep{qui14} is a five-planet system where the farthest planet from the star, Kepler 
186f, is located at 0.432($+0.171,-0.053$) AU \citep{tor15} and within the habitable zone of the host star. The nearest planet from the star, Kepler 186b, is at
0.0378 AU and orbits the star every 3.88 days. Another such planetary system is Gliese 581 \citep{may09}, which 
is known to host three planets along with the two others that are not yet confirmed. The three known planets, Gliese 581b,c, e, orbit the star within 0.07 AU.
Also, many binary star systems have been discovered which are known to host multiple circumprimary planets in S-type orbits, for example 55 Cancri b,c,d,e,f \citep{fis08}. With an addition of the third planet in the Kepler 47 system \citep{oro15}, it became the first multi-planet circumbinary system and opened a new chapter for us to understand the planet formation processes and the dynamical compactness of the planetary orbits.

The GJ 832 planetary system \citep{wit14} is another multi-planet system which is currently known to host two planets around an M dwarf star, and is located at a relatively close distance of 16.1 light years from Earth. GJ 832c (inner planet) orbits its star at a distance 0.163$\pm$ 0.006 AU away and is potentially a rocky planet with a mass 5.4$\pm$0.95 M$_\oplus$. This planet is located in the inner boundary of the habitable zone, but it is not expected to be habitable mostly due to its close proximity to the star and its possibility of having dense atmosphere \citep{wit14}. Orbiting the same star, a distant planet GJ 832b (outer planet) was discovered by \cite{bai09}; which is a long-period, 3657$\pm$104 days giant planet at 3.56$\pm$0.28 AU, having a mass (\emph{msini}) of 0.64$\pm$0.06 M$_J$. The planets orbit a main sequence dwarf star, GJ 832, of a spectral type M1.5V \citep{jen06}, with a mass of 0.45$\pm$0.05 M$_\odot$ \citep{bon13}, and a temperature of 3472 K \citep{cas08}. This is fairly an old system. \cite{gui16} have estimated the age of the star to be 6$\pm$1.5 Gyr by using their X-ray Activity-Age relationship. 
We have calculated the size of the stellar habitable zone by using the formula provided by \cite{kop13}. Then, the orbital stability of additional planets having mass $\le$15 M$_\oplus$ is investigated within and outside the boundaries of this habitable zone.

The main goal of this article is to explore the gravitational effect of the outer planet on the orbital stability of the inner planet as well as on planets with mass $\le$15 M$_\oplus$ injected between the two known planets. In addition, the long-term stability and orbital configurations of the inner and the injected planets, with concentration on the time evolution of their semimajor axis ($a_{pl}$), eccentricity ($e_{pl}$) and inclination ($i_{pl}$), is studied in the $a_{pl}$ - $e_{pl}$ and $a_{pl}$ - $i_{pl}$ phase-spaces, and time-series 2-D plots.

We also used the integrated data from the time evolution of orbital parameters to generate synthetic radial velocity (RV) curves of the known planets as well as the added planets with masses varying from 1 M$_\oplus$ to 15 M$_\oplus$ in the system. Moreover, based on the maximum amplitude of the RV curve obtained from the observation of the inner planet, the approximate mass and distance from the star for the potential Earth-mass planet were computed using the RV signature of the Keplerian motion.

This paper is outlined as follows: In Section 2, we describe our numerical simulations; the results are presented and discussed in Section 3; and the paper is concluded with a brief summary of our main results in Section 4.

\section{Numerical Simulations}

The best-fit orbital parameters of the GJ 832 system as obtained from the original discovery papers \citep{wit14,bai09} are given in Table \ref{tab:1} and for the cases when the parameters are unknown, they are either set close to zero or set to a range with a fixed step-size. Both of the known planets in the GJ 832 system were detected by the RV technique from which the orbital parameters were extracted by using the best-fit Least-Squares Keplerian Orbital Solutions. We used these parameters as the initial conditions for starting our numerical simulations. The best-fit orbital solutions include uncertainties in their respective parameters. For example, the semimajor axes of the inner and outer planets with uncertainty are 0.163$\pm$0.006 AU and 3.56$\pm$0.28 AU, and the masses (\emph{msini}) are 5.4$\pm$0.95 M$_\oplus$ and 216$\pm$28 M$_\oplus$, respectively. In some of our phase-space analysis, we have considered the maximum uncertainty values, specially in the planets' eccentricities, semimajor axes and masses. This has allowed us to see the gravitational effects on the long-term orbital stability of the inner planet as well as the injected planet for some extreme values of their orbital paramters.

We have considered the motion of the planets of masses, $m_{pl}$ around the central star in the general elliptical as well as the circular cases. To calculate the planets' and the star's initial conditions (ICs) in terms of their position and velocity and start the integration processes, we used their best-fit orbital elements: semi major axis ($a$), eccentricity ($e$), inclination ($i$), argument of periapsis ($\omega$), ascending node ($\Omega$) and mean anomaly ($M$), which were obtained from RV measurements \citep{wit14}. The initial orbital inclination of the inner and the injected Earth-mass planets are taken relative to the orbital plane of the star and the outer planet. Thus, any inclination we mention during our investigation is relative to the star - outer planet plane. To check the stability of the system for other limiting cases, we have used the upper limit of the best-fit orbital elements. Note that, in this paper, the \emph{stability} of the system is defined in terms of its \emph{lifetime}, which is based on the time a planet survives the total simulation time, collides with other bodies or gets ejected from the system during the orbital integration period. The \emph{stability} is also studied in term of the \emph{maximum eccentricity}, $e_{max}$ attained by the planetary orbits during their evolution processes. A system is considered \emph{stable} when the integrated bodies survive the total simulation time and their $e_{max}$ deviate least from the initial value; else, the system is considered unstable.

Using the orbital integration package MERCURY \citep{cha97,cha99}, the built-in Hybrid algorithm was used to integrate the orbits of the system in astro-centric coordinates. MERCURY was effective in monitoring the ejection or collision of the inner and the injected planets due to a close encounter with the star or the outer planet. While integrating the orbits, a time step of $\epsilon = 10^{-3}$ year/step was considered to obtain high precision data and minimize the error accumulation. The change in total energy and total angular momentum was calculated at each time step which fell within the range of 10$^{-16}$ to 10$^{-13}$, respectively during the total integration period of 10 Myr, and the range of 10$^{-10}$ to 10$^{-12}$ during 1 Gyr. The data sampling (DSP) was done per day and per year for shorter integration periods and at every 100 kyr for billion years integration period. The \emph{lifetime} maps and the maximum eccentricity ($e_{max}$) maps are generated for multiple (up to 14,400) initial conditions in $a_{pl}$, $e_{pl}$ and $i_{pl}$ phase-spaces, and they are simulated for 10 Myr. The billion years simulations are performed for low resolution phase-space maps and for some selected single initial conditions.

\section{Results and Discussion}
\subsection{Dynamics of GJ 832c}

The inner planet, GJ 832c, does not have well constrained orbital parameters (Table \ref{tab:1}), including its orbital inclination ($i_{pl}$) and the longitude of ascending node ($\Omega_{pl}$) which do not have the best-fit values. Therefore, to set up our initial conditions for the simulations, we have set $\Omega_{pl}$ to (10$^{-5}$)$^{o}$ and $i_{pl}$ to a value between near co-planar (10$^{-5}$)$^{o}$ and 90$^{o}$. Other orbital parameters are set at their best-fit nominal values given in Table \ref{tab:1}. Similarly, to set up the initial conditions for the outer planet, GJ 832b, its best-fit nominal values are considered when available; otherwise they are set close to zero. We have also performed simulations where the initial conditions are set at upper values of the uncertainty limit. The lower uncertainty values of the orbital parameters would have less effect on the stability of the phase-spaces between the known planets, which is our major region-of-interest for an additional planet; hence, no simulations are run for such cases. The $\omega_{pl}$, $\Omega_{pl}$, and $M_{pl}$ are considered slow moving angles and have the least effect on the orbital stability. The influence of these angles are more significant in the resonant angle studies which determines the libration or circulation of the phase angles \citep{mur99}. Also, see \cite{sat14} for the analysis of these angles in the study of the chaotic dynamics of the planet in HD 196885 binary system.

To investigate the orbital inclination of the inner planet, its orbits are integrated with 14,400 initial conditions (ICs) in varying $i_{pl}$ - $a_{pl}$ phase-space, and 8,000 ICs in $e_{pl}$ - $a_{pl}$ phase-space. To perform our simulations, we only considered prograde orbits where $i_{pl}$ is sampled from (10$^{-5}$)$^{o}$ to 90$^{o}$ with a step size of 0.5$^{o}$, $a_{pl}$ is sampled from 0.1 to 4.0 AU with the step size of 0.05 AU, and finally $e_{pl}$ is sampled from 10$^{-5}$ to 1.0 with a step size of 0.05. Then, within a block of [($i_{pl}$ or $e_{pl}$), $a_{pl}$], each of the ICs mentioned above are set to evolve for up to 10 Myr and 1 Gyr. During this integration period, the close encounters, ejections, and collisions between the planets and the host star are allowed to occur, which marks the stoppage of the integration processes for those ICs. If the integrated orbit survives the total simulation time, then we consider it to be a stable orbit. However, in some cases when the integrated bodies eject or collide during close encounters, displaying instability of the system, we note the time of such events and use that time to create a global dynamical \emph{lifetime} map that displays dynamically stable or unstable regions.

\subsubsection{Analysis of the \emph{lifetime map}: GJ 832c}

\begin{figure}[!ht]
\centering
\includegraphics[width=.8\linewidth]{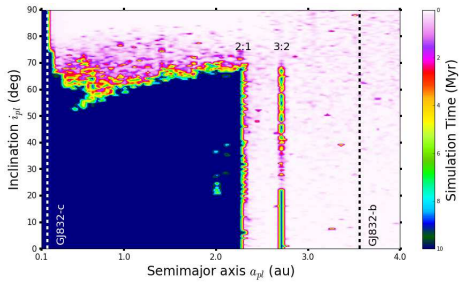}
\caption{A global dynamical \emph{lifetime} map of the inner planet, GJ 832c, in varying $i_{pl}$ and $a_{pl}$ phase-space, simulated for 10 Myr. The map represents the time evolution of the orbital elements with 14,000 initial conditions. The survival time (maximum 10 myr) is plotted on the color coded z-axis. The index in the color bar indicates the survival time, where the lighter colors represent the instability (ejection or collision) and the dark-blue color represents the stability (survival) up to the integration period. Hence, the dark blue-green region in the map is the dynamically stable zone. The vertical dashed lines at 0.16 AU and 3.56 AU represent the best-fit semi-major axis of inner and outer planets. The vertical islands at 2.25 and 2.7 AU are in 2:1 and 3:2 orbital resonance with the outer planet. These resonances are more prominent in the phase-space map for $e_{max}$ (Fig. \ref{fig:2}).}
\label{fig:1}
\end{figure}

To explore the dynamics of the inner planet, its \emph{lifetime} map (Fig. \ref{fig:1}) is created for multi ICs in $i_{pl}$ and $a_{pl}$ phase-space. Each grid point of the map represents one IC. The color coding of each IC gives the survival time of the planet in the respective simulation; however, when a smaller number of data points are available, the interpolation method is used to construct data points in the neighborhood of the known points. The color codes are in the z-axis with index given in the right-hand color-bar. The dark blue color in the map indicates the survival of the planet for 10 Myr, which in this case corresponds to the total simulation time. Lighter colors (from blue to white in our color-bar) represent unstable dynamical configurations, indicating that the planet was ejected from the system or collided with the star or the outer planet in less than 10 Myr. The vertical dashed lines in the figure, labeled as GJ 832c and GJ 832b, represent the best-fit semimajor axis of the planets. At the best-fit location, GJ 832c remains in a stable orbit for the total simulation period with an orbital inclination as high as 70$^\circ$. With longer simulation times ($>$ 10 Myr), this inclination regime may significantly be reduced which we explore in the following sections. On the contrary, during the 10 Myr integration period the inner planet with an initial orbital inclination larger than 60$^\circ$-70$^\circ$, either collided with the star or was ejected from the system. This is because when the perturber (GJ832b) has an eccentric orbit, the inner planet's orbital inclination and eccentricity may reach extremely large values due to the Lidov-Kozai \citep{lid62,koz62} effect \citep{sat14,nao13,for00}. This effect, especially when the $i_{pl}$ crosses certain limit, eventually leads the system towards instability which is observed in Fig. \ref{fig:1}.


\subsubsection{Maximum Eccentricity in $a_{pl}$, $e_{pl}$ and $i_{pl}$ Phase-Spaces: GJ 832c}

\begin{figure}[!ht]
\centering
\hspace*{\fill}
\includegraphics[width=1.0\linewidth]{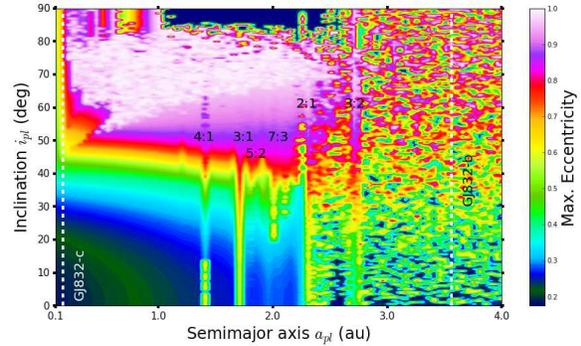}
\hspace*{\fill}%

\caption{Maximum eccentricity, $e_{max}$ map of the inner planet, GJ 832c, in $i_{pl}$ and $a_{pl}$ phase-space, simulated for 10 Myr. The map represents the time evolution of the planet's eccentricity for 14,400 initial conditions. The index in the color bar indicates the $e_{max}$ reached by the planet during the total simulation time, which also includes the cases when the planet suffers an ejection or collision (especially when $e_{pl}$ reaches a value greater than 0.7). The dark blue-green color represents the best-fit $e_{pl}$ value from the observational data ($\sim$0.18) and other light colors represent the $e_{max}$ for the respective choices of initial conditions in $i_{pl}$ and $a_{pl}$. The vertical white dashed lines denote the best-fit semi-major axis of the planets. The observed resonances at 1.4, 1.7, ,1.9, 2.0, 2.25 and 2.7 AU are in 4:1, 3:1, 5:2, 7:3, 2:1, and 3:2 orbital resonance with the outer planet. Similar resonances (3:1, 5:2, 7:3 and 2:1) are observed in the Solar System due to Jupiter's influence in the Asteroid Belt, called Kirkwood gaps.}
\label{fig:2}
\end{figure}

The maximum eccentricity ($e_{max}$) map is shown in Fig. \ref{fig:2} for the same phase-space, $i_{pl}$ - $a_{pl}$, as in Fig. \ref{fig:1}. The color coded z-axis  in the map represents the maximum orbital eccentricity attained by the inner planet during the total integration period. This $e_{max}$ map was obtained by integrating 14,400 ICs for 10 Myr and recording the maximum eccentricity for each sate during their time evolution. Therefore, the $e_{max}$ value for each of the states may or may not be equal to the final eccentricity attained at the end of the simulation. In addition, a major fraction of these maximum values, especially in the unstable regions and around its borders with the stable regions, are expected to rise with significantly increased simulation time (see Section 3.1.3 for 1 Gyr simulations).

The $e_{max}$ map in Fig. \ref{fig:2} shows the maximum eccentricity values that evolved from their nominal values during 10 Myr simulation, and for different ICs of the inner planet's orbital inclination which is plotted along the \emph{y}-axis. The planet's $e_{max}$ value increases with increasing $i_{pl}$ along the best-fit $a_{pl}$ value (0.163 AU, white dashed line), observed in the figure in various color schemes (blue-green-orange-pink regions). The planet's $e_{max}$ remains less than 0.5 in the regions where the $i_{pl}$ is less than 45$^\circ$. For higher $i_{pl}$ values, $e_{pl}$ quickly increases to 1. Thus, the map suggests that the likely $i_{pl}$ value is less than 45$^\circ$. Note that the blue stable region for $i_{pl}$$>$80$^\circ$ and 1$<$$a_{pl}$$<$2 indicates $e_{max}$$<$0.1. This is because the planet attained maximum eccentricity ($\sim$1) in a time shorter than our \emph{data sampling period}. Hence, the simulation algorithm recorded the initial eccentricity value instead. Furthermore, it is unlikely for the planet to have $i_{pl}$ greater than 45$^\circ$ and still maintain a stable orbit because the observed trend in the map shows the $e_{max}$ quickly increases to 0.8 and 0.9. Therefore, with increasing $i_{pl}$, the $e_{max}$ value is only expected to rise for longer simulation time. Hence, the blue-green region is the most dynamically stable zone for the inner planet. The dynamics of the region between the two planets, especially GJ 832c, changes with an additional bodies injected between them, which we explore in Sec. 3.2.

\begin{figure}[!ht]
\centering
\hspace*{\fill}
\includegraphics[width=1.0\linewidth]{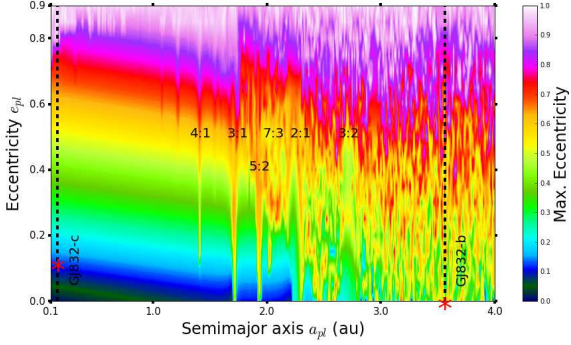}
\hspace*{\fill}%

\caption{Maximum eccentricity, $e_{max}$ map of the inner planet in $e_{pl}$ - $a_{pl}$ phase-space, simulated for 10 Myr. The red asterisks denote the best-fit eccentricity value of the inner and middle planets, which shows no deviation from the nominal eccentricity values during the total integration period. The vertical resonance at 1.4, 1.7, ,1.9, 2.0, 2.25 and 2.7 AU are in 4:1, 3:1, 5:2, 7:3, 2:1, and 3:2 orbital resonance with the outer planet.}
\label{fig:3}
\end{figure}

The inner planet's $e_{max}$ data shown in $e_{pl}$ and $a_{pl}$ phase-space (Fig. \ref{fig:3}) also reveals vertical structures corresponding to the location of mean motion resonances with the outer planet and it complements, in terms of stability and instability regions, the $e_{max}$ map in Fig. \ref{fig:2} plotted for different phase-space. The y-axis displays different initial conditions for $e_{pl}$, while the color coded z-axis shows the $e_{max}$ attained by the respective initial conditions. The $e_{max}$ value demonstrates lesser variation from its nominal value, seen in the horizontal blue region along the red asterisk at 0.163 AU in Fig. \ref{fig:3}. The best-fit $e_{pl}$ value is along the best-fit $a_{pl}$ values (vertical dashed lines) for both planets and are denoted by the red asterisks. When the $e_{pl}$ is set at 0.31, the upper uncertainty value, it does not show any significant deviation during the full integration period, suggesting that the uncertainty in the $e_{pl}$ cannot be constrained any further from this analysis.


\subsubsection{Effects of Uncertainty Limits in the Stability of the System}

To check how the uncertainty of the orbital elements in Table \ref{tab:1} affects our conclusions on the stability of the system, we integrated the orbits with the upper uncertainty values of the planets' mass, eccentricity, mean anomaly, and argument of periapsis. The upper limit of the semimajor axis is chosen for the inner planet and the lower limit for the outer planet. This allows us to see how the stable phase-space that are observed in the above sections changes when the outer giant planet, for its upper limit in mass, is placed closer to the inner planet.

\begin{figure}[!ht]
\centering
\hspace*{\fill}
\includegraphics[width=1.0\linewidth]{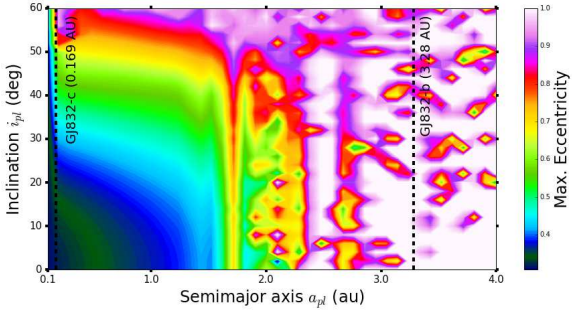}
\hspace*{\fill}
\caption{Maximum eccentricity, $e_{max}$ map of the inner planet, GJ 832c, in $i_{pl}$ and $a_{pl}$ phase-space, simulated for 1 Gyr and 1,400 initial conditions. The semimajor axis of the inner planet is set at 0.169 AU (upper uncertainty limit of its best-fit value) and the outer planet is set at 3.28 AU (lower uncertainty limit of its best-fit value). Both of the planetary masses and the other orbital parameters are set at their upper uncertainty limit as well. The index in the color bar indicates the $e_{max}$ reached by the planet during the total simulation time, which also includes the cases when the planet suffers an ejection or collision. The instability region has shifted inward up to $\sim$1.5 AU.}
\label{fig:4}
\end{figure}

The maximum eccentricity reached by the inner planet after 1 Gyr of orbital evolution is plotted in the $e_{max}$ map shown in Fig. \ref{fig:4}. Compared to the previous maps, this map has lower resolution ($\sim$1,400 ICs in \emph{x} and \emph{y} axes). Nevertheless, the $e_{max}$ for the semimajor axes between 0.169 AU and 1.5 AU shows no significant deviation from the assumed initial $e_{pl}$. But, beyond 1.5 AU, the map shows rise in the instability islands where the $e_{max}$ is found to reach values close to 1. For most of the cases when the inner planet's eccentricity reached more than 0.5, it was ejected from the system or collided with the outer planet.

Comparing the 10 Myr simulation map (Fig. \ref{fig:2}) with the 1 Gyr simulation map (Fig. \ref{fig:4}), we see that the planet's orbital inclination between 40$^\circ$ and 50$^\circ$ shows the eccentricity variation from 0.3 to 0.4 and 0.4 to 0.5, respectively. That is, 0.1 eccentricity variation is observed in the same inclination regime for the increased simulation time. The unstable region, as expected, has extended further inward from the location of the outer planet which is set at 3.28 AU, the lower limit of its semimajor axis. Also, the observed resonance structures in Fig. \ref{fig:2} have started to diffuse within the phase-space which is primarily due to the lower resolution map and shifting of the high mass outer planet closer towards the inner planet.

For the lower limit in the uncertainty values of the orbital elements and the planetary masses, we expect the stability region to remain unaltered, if not widen a little, when compared to the map shown in Fig. \ref{fig:2}, where the nominal values of the best-fit orbital parameters are used to integrate the system.


\subsection{Dynamics of an Additional Planet}

The known planetary configuration in this system shows a super Earth orbiting at the close proximity, 0.163$\pm$ 0.006 AU from the host star, while a gas giant orbits distantly at 3.56$\pm$ 0.28 AU (for reference: Mercury orbits the Sun at 0.39 AU and Jupiter at 5.2 AU). Therefore, the existence of other Earth-mass planet(s) (could be bigger or smaller than Earth) between the inner and the outer planets is a plausible scenario. As observed in Figs. \ref{fig:1} - \ref{fig:4}, a planet with mass equal to 5.4 M$_\oplus$ and orbiting between 0.1 AU to 2.0 AU is dynamically stable for a wide range of initial orbital eccentricities and inclinations. However, it is also important to note that even inside 2.0 AU unstable orbits also exist as for example those near mean motion resonances. To observe how the stability of the system changes with an additional bodies, we injected a third planet (\emph{middle} planet from herein), having 1 M$_\oplus$ and studied its orbital dynamics in the similar phase-spaces that are discussed in Sec. 3.1.

Orbital parameters for the middle planet are chosen based on the stability zone observed in Figs. \ref{fig:1} - \ref{fig:4}. For each initial configuration, $i_{pl}$ is varied from (10$^{-5}$)$^\circ$ - 90$^\circ$, $a_{pl}$ from 0.1 to 4 AU, and  $e_{pl}$, $\Omega_{pl}$ and $\omega_{pl}$ are set close to zero while $M_{pl}$ is randomly chosen between (10$^{-5}$)$^\circ$ - 360$^\circ$. The mass is first set at 1 M$_\oplus$, and later raised up to 15 M$_\oplus$ to observe the orbital variations within the stability zone. The nominal best-fit values of the orbital parameters are used for the inner and outer planets while integrating the middle planet. Later, in the second case while studying the phase-space of the middle planet, the upper limit of the known planets' orbital parameters, including their masses, are considered and the system is integrated for up to 1 Gyr.

\subsubsection{Lifetime and Maximum Eccentricity Maps: the Middle Planet}

\begin{figure}[!ht]
\centering
\includegraphics[width=1.0\linewidth]{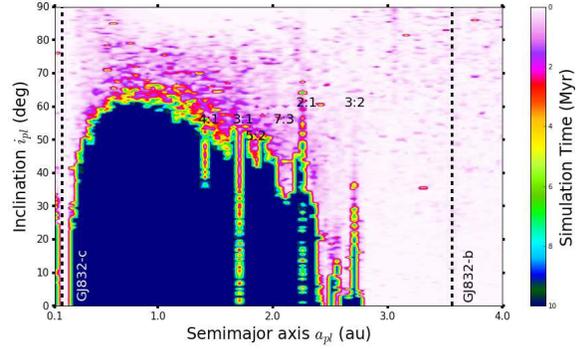}
\caption{A global dynamical \emph{lifetime} map of an Earth-mass middle planet in GJ 832 system injected in the dynamically stable zone (observed in Fig. \ref{fig:2}) between the inner and the outer planet. The map represents the survival time of the middle planet for 14,400 ICs in $i_{pl}$ - $a_{pl}$ phase-space and simulated for 10 Myr. The index in the color bar indicates the survival time, where the lighter colors represent the instability (ejection or collision) and the dark-blue color represents the stability (survival) up to the integration period. The vertical dashed lines are the locations of the best-fit semimajor axis of the two known planets. The dark-blue region indicates the stable orbital configuration for the Earth-mass planet.}
\label{fig:5}
\end{figure}

The \emph{lifetime} map and the $e_{max}$ map (Figs. \ref{fig:5} and \ref{fig:6}) of the injected middle planet with 1 M$_\oplus$, are generated from its \emph{survival time} in the orbits and the maximum eccentricity attained by the orbits during the total integrating period of 10 Myr, respectively. These maps indicate a wide stability region in the $i_{pl}$ - $a_{pl}$ phase-space. The blue region in the \emph{lifetime} map, and the blue-green region in the $e_{max}$ map, extends from  0.2 AU to 2.2 AU along the semimajor axis and 0$^\circ$ to 40$^\circ$ (on the average) along the inclination axis. The map shows sharp changes in the $e_{max}$ values along the  $\sim 40^\circ$ $i_{pl}$ value, seen along the horizontal strip of cyan-blue marker. This is the Lidov-Kozai \citep{lid62,koz62} resonance above which the planet is orbitally unstable. The $e_{max}$ values in the region of phase-space which is below $40^\circ$ is less than 0.15 and all the orbits survive the total simulation time. Outside this stable zone, the middle planet's eccentricity was forced to 0.4 or higher in most of the cases which caused it to either collide with the inner planet or get ejected from the system. In either case, the planet lost its orbital stability.


\subsubsection{Long-term Orbital Stability: the Middle Planet}

The \emph{lifetime} and $e_{max}$ maps discussed in Sec. 3.2.1 have shown a region where the 1 M$_\oplus$ planet can maintain stable orbit between the two known orbits, and for up to 10 Myr. For similar phase-space, we then integrated the system for 1 Gyr and raised the mass of the middle planet to 15 M$_\oplus$. The primary reason we chose 15 M$_\oplus$ planet is because, if the system is stable for this mass, it is more likely to be stable for lower mass planets that could exist between the inner and outer planets. The other reason is discussed in Sec. 3.3 on how the maximum amplitude of the RV signal is produced by a 15 M$_\oplus$ planet when the \emph{sin(i)} is considered to be maximum (\emph{i} $\sim 90^\circ$). In addition, we set the masses, eccentricity and argument of periapsis of the inner and outer planet to their upper uncertainty limit as given in Table 1. Also, the orbit of the inner planet is moved outward to its maximum value (0.169 AU) and the outer planet is moved inward to its minimum value (3.28 AU) to minimize the space between them and see how that affects the dynamics of the middle planet.

Figure \ref{fig:7} shows the maximum eccentricity of the 15 M$_\oplus$ middle planet in $i_{pl}$ - $a_{pl}$ phase-space after 1 Gyr orbital evolution with $\sim$1,400 initial conditions in vertical and horizontal axes. The semimajor axes of the inner and outer planets are denoted by the black dashed lines. The color bar indicates the maximum eccentricity attained by the middle planet during the total integration period. The dark blue-green regions suggests a potentially stable zone where the e$_{max}$ is less than 0.2, while the other brighter colors represents potentially unstable zones where the $e_{max}$ reaches close to 1. The eccentricity is found to deviate least from its initial value of 10$^{-5}$ around the 1 AU marker and remains less than 0.1 for $i_{pl}$ up to $\sim$30$^\circ$. The vertical structure observed close to 1.35 AU corresponds to the location of the near 4:1 mean motion resonance with the outer planet and in our case all selected orbits around this region are unstable ones.

\begin{figure}[!ht]
\centering
\hspace*{\fill}
\includegraphics[width=1.0\linewidth]{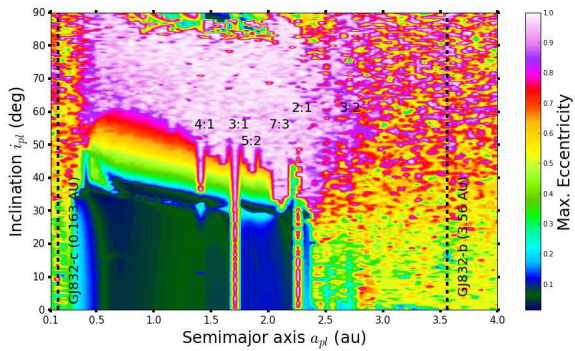}
\hspace*{\fill}%
\caption{Maximum eccentricity ($e_{max}$) map of the Earth-mass planet injected between the inner and outer planets, in the $i_{pl}$ and $a_{pl}$ phase-space and simulated for 10 Myr with 14,400 ICs. The index in the color bar indicates the $e_{max}$ the orbit evolved into, starting from 10$^{-5}$ during the total simulation time. The $e_{pl}$ varies less from 0.2 to 2.2 AU and below 40$^\circ$ inclination regime. In any other cases, the planet suffered an ejection or collision with the inner or outer planets, especially when $e_{pl}$ reaches a value greater than 0.5. The dark blue-green color represents the best-fit $e_{pl}$ parameter from observation (0.18) and other light color represent the $e_{max}$ value the planet attained for the respective choices of initial conditions in $i_{pl}$ and $a_{pl}$. The vertical dashed line are the best-fit semimajor axis of the two known planets, and the resonances due to the outer planet are similar to that observed in Fig. \ref{fig:2}.}
\label{fig:6}
\end{figure}

The stability regions get narrower when compared to the phase-space map generated for the 1 M$_\oplus$ planet observed in Fig. \ref{fig:6}. The inclination regime has reduced to $\sim$15$^\circ$ for the $a_{pl}$ between 0.5 to 0.75 AU. Also, the overall $i_{pl}$ regime for the middle planet is less than $\sim$30$^\circ$ and extends roughly from 0.5 to 2.0 AU. The map suggests that the best possible location for the middle planet would be in the proximity of the 1.0 AU marker where the eccentricity undergoes least deviation from its initial value.

The phase-space around the GJ 832c at 0.169 AU (black dashed line) has undergone the least eccentricity variation compared to the e$_{max}$ map in Fig. \ref{fig:6}. This does not mean that the injected 15 M$_\oplus$ planet is stable in that region. The maximum eccentricity has remained very close to $\sim$0.1 (represented by the blue color in the map) because the high-mass injected planet collides with the low-mass inner planet (5.4 M$_\oplus$), changing the orbital configuration of the latter and making it dynamically unstable. During this collision, the maximum eccentricity attained by the 15 M$_\oplus$ planet deviates least from its nominal value, and this is what we have plotted in the map. Also, in the higher inclination regimes, the inner planet displays more variation in its eccentricity compared to the middle planet, which we discuss in the next section.


\subsubsection{Eccentricity and Inclination Time Series of the Planets}

\begin{figure}[!ht]
\centering
\hspace*{\fill}
\includegraphics[width=1.0\linewidth]{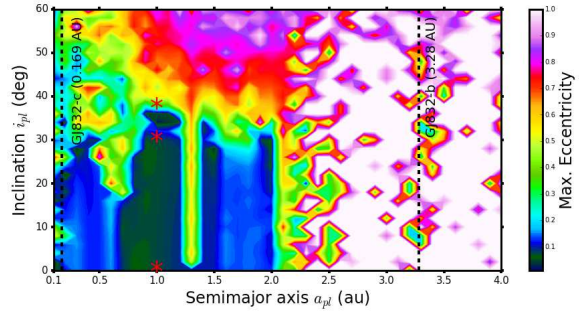}
\hspace*{\fill}%
\caption{Maximum eccentricity ($e_{max}$) map of a 15 Earth-mass planet injected between the inner and outer planets, in the $i_{pl}$ and $a_{pl}$ phase-space and simulated for 1 Gyr with 1,400 ICs. The index in the color bar indicates the $e_{max}$ attained by the orbit during its evolution starting from an initial value of $e_{pl}$ = 10$^{-5}$. The $e_{pl}$ varies less from 0.2 to 2 AU and below 30$^\circ$, with a few exception of instability islands. In any other cases, the planet suffered an ejection or collision with the inner or outer planets, especially when $e_{pl}$ reaches a value greater than 0.5. The dark blue-green color represents the best-fit $e_{pl}$ parameter from observation (0.18) and other light colors represent the $e_{max}$ value the planet attained for the respective choices of initial conditions in $i_{pl}$ and $a_{pl}$. The vertical dashed line are the upper and lower uncertainty limits of the best-fit semi-major axes of the inner and the outer planets, respectively. The masses of the known planets are set at their upper uncertainty limit. The three specific locations, denoted by the red asterisks, are explored for a long-term orbital evolution in Fig. \ref{fig:8} in an attempt to constrain the injected planet's $i_{pl}$ and $e_{pl}$ and observe the orbital evolution in time series 2-D plots.}
\label{fig:7}
\end{figure}

\begin{figure}[!ht]
\centering
\includegraphics[width=.6\linewidth]{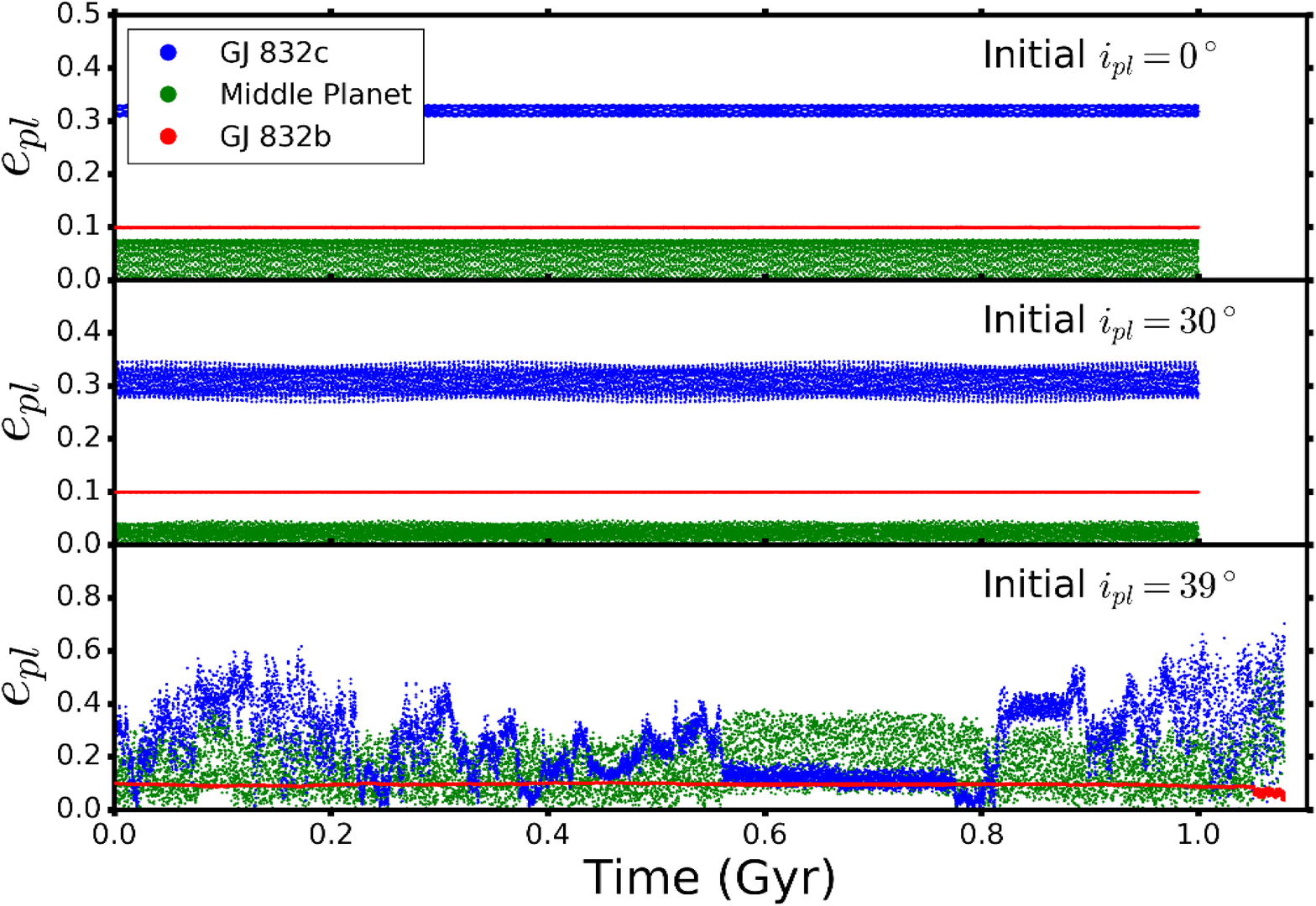}
\includegraphics[width=.6\linewidth]{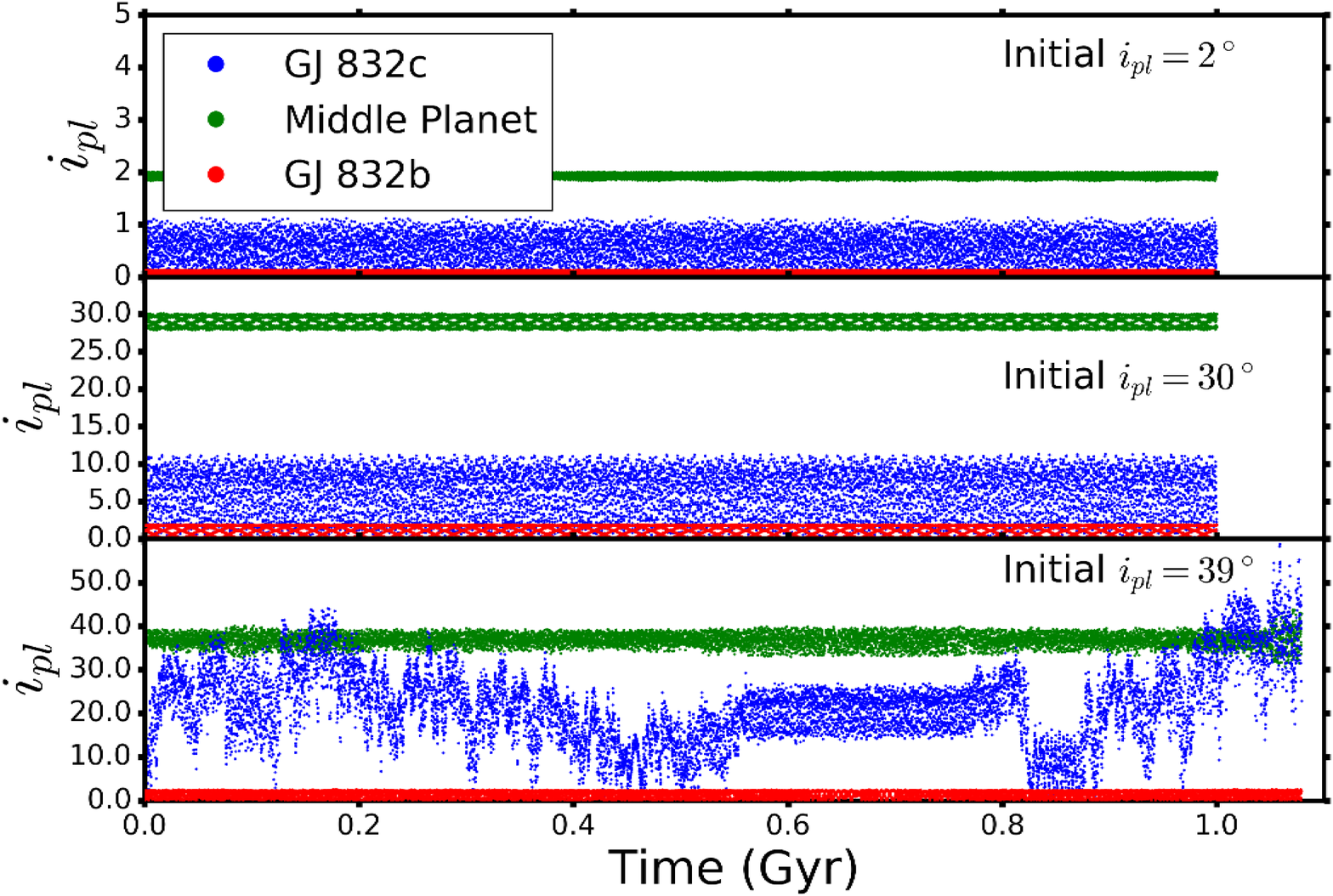}
\caption{[Top] The eccentricity time-series for the planets in GJ 832 system when the middle planet's $a_{pl}$ is set at 1 AU, mass is set at 15 M$_\oplus$ and $i_{pl}$ at 0$^\circ$ (top panel), 30$^\circ$ (middle panel), and 39$^\circ$ (bottom panel) orbital inclination relative to the stellar plane. The upper uncertainty limit of the best-fit values of $e_{pl}$ were used for the inner and outer planets, and the initial $e_{pl}$ for the middle planet was set close to zero (10$^{-5}$). For $i_{pl}$ = (10$^{-5}$)$^\circ$, the eccentricity evolution is smooth with minimal deviation from its initial value for the inner and the outer planets. The amplitude of $e_{pl}$ oscillations slowly increases as the $i_{pl}$ is raised higher, and is highly noticeable for the middle and inner planets when $i_{pl}$ = 30$^\circ$. The orbits of the middle and the inner planets display chaotic orbits for higher initial $i_{pl}$ values, and finally become unstable at $i_{pl}$ = 39$^\circ$. The middle planet, which is set at 15 M$_\oplus$, has more gravitational perturbation on the inner planet which has lesser mass (6.35 M$_\oplus$). Therefore, the eccentricity of the middle orbit remains less perturbed and the inner, more chaotic. [Bottom] Stable and chaotic orbits are observed in the inclination time-series for the similar orbital configuration for the planets in the GJ 832 system. The $i_{pl}$ of the inner planet displays more chaotic behavior for higher $i_{o}$ values of the middle planet. The $e_{pl}$ and $i_{pl}$ evolution for $i_{pl}$ = 39$^\circ$ case shows the anti-correlation between them. [Orbital simulation for the two cases, $i_{pl}$ = 0$^\circ$ and 30$^\circ$, are stopped at 1 Gyr primarily because no significant dynamical variations were observed in the orbital parameters, and because of the limited computational time.]}
\label{fig:8}
\end{figure}

To confirm that the dark blue-green regions as seen in Fig. \ref{fig:7}, which we claim to be an orbitally stable zone for the middle planet, continues to remain in long-term stable orbits and to observe a time series evolution of $e_{pl}$, we picked a few initial $i_{pl}$ points along the 1 AU mark (see Fig. \ref{fig:7}, red asterisks) and re-integrated for the selected ICs. The other orbital parameters that are used to set up the ICs are similar to the ones we discussed in the previous Sec. 3.2.2. The masses of the known planets are set at their upper uncertainty limit of the best-fit values given in Table 1, and the injected middle planet's mass is set at 15 M$_\oplus$. The time evolution of the middle planet's eccentricity for three different orbital orientations (0$^\circ$, 30$^\circ$ and 39$^\circ$) are given in Fig. \ref{fig:8}. The initial $e_{pl}$ and $a_{pl}$ was set to 10$^{-5}$ and 1 AU, respectively, and the planet was allowed to evolve in the gravitational influence of the two known planets and the star.

Time series $e_{pl}$ evolution (Fig. \ref{fig:8}, top plots): For $i_{pl}$ = 0$^\circ$ (Fig. \ref{fig:8}, top panel), the amplitude of the eccentricity oscillations remains near the initial values for the known planets, and varies between 0 and 0.08 for the middle planet. The observed variation in the eccentricity time series is less significant until the initial inclination ($i_{o}$) is set above 30$^\circ$. For $i_{o}$ = 30$^\circ$, the eccentricity time-series of the inner planet starts to display larger amplitude oscillations (0 to 0.1); however, no violent end is observed during the billion year simulation time. Finally, for $i_{o}$ = 39$^\circ$ the inner and the middle planet's eccentricity time-series vary significantly in their amplitudes, displays chaotic evolution, and eventually evolves closer to 1. Then, the middle planet collides with the inner planet at around 1.081 Gyr leading to the system's instability.

Similarly, we looked at $i_{pl}$ evolution (Fig. \ref{fig:8}, bottom plots) of all three planets for three different cases when the middle planet's initial inclination ($i_{o}$) is set at 2$^\circ$, 30$^\circ$ and 39$^\circ$, and eccentricity at 10$^{-5}$. The amplitude of the $i_{pl}$ oscillations start to rise significantly for higher $i_{o}$. For example, when $i_{o}$ is set at 30$^\circ$ and 39$^\circ$, the inner planet's inclination reaches 10$^\circ$ and 60$^\circ$, respectively, from its near co-planar orbit and displays more chaotic orbits for the latter case. Recent studies \citep{bar16} of such similar chaotic orbits, with large variations in eccentricity and inclination, have shown stable orbits for up to 10 Gyr.

The Lidov-Kozai resonance occurs at 39.2$^\circ$ (see \cite{lid62, koz62, inn97}), beyond which the anti-correlation between the $e_{pl}$ and $i_{pl}$ excites the orbits into high eccentricities, significantly reducing the periastron distance and leading to a collisional path (for detailed Kozai resonance analysis, see \cite{sat14}). Also, a recent study of the planets in circumbinary orbits of $\alpha$ Centauri AB by \cite{qua16} shows how the Lidov-Kozai resonance beyond 40$^\circ$ limits the stability region. Therefore, we believe that the maximum inclination for the planets in the GJ 832 system, which are orbiting interior to the outer planet, is less than the critical angle of 39.2$^\circ$.

The above discussion is for the case when initial eccentricity ($e_o$) of the middle planet is set to near zero. Our separate tests show that the system looses its stability due to a collision between the inner and middle planets shortly after 700 Kyr of integration time when $e_o$ is set above 0.4. For these sets of simulations, the initial orbital inclination was set close to zero in all cases, and the other orbital parameters remained the same as discussed above.

\subsubsection{Orbital Resonances in Presence of the Middle Planet}

Fig. \ref{fig:6} does not exhibit any orbital resonant structures corresponding to the expected locations of MMRs, specially in the outer regions of the inner planet's best fit location at 0.163 AU. We expect such structures to emerge with significantly longer simulation time and in a higher resolution map. In addition, the prominent vertical structures are observed beyond 1.0 AU which correspond to the unstable regions near the locations of MMRs with the outer planet. These resonances at the 1.40 AU, 1.70 AU, 1.93 AU, 2.00 AU, 2.25 AU and 2.70 AU are due to the 4:1, 3:1, 5:2, 7:3, 2:1 and 3:2 MMRs, respectively. Similar resonance structures are observed in Solar System's asteroid belt, where the \emph{Kirkwood} gaps are in the 3:1, 5:2, 7:3, and 2:1 resonance with Jupiter \citep{moo95}. However, not all resonances are unstable. For example, Jupiter's moons (Io, Europa and Ganymede) form a resonant system with 1:2:3 orbital resonance and maintain stable orbits.

The vertical structures between 1.5 AU to 2.5 AU corresponding to the location of mean motion resonances fade away and shift inwards when the outer planet's location is moved to 3.28 AU, as shown in Fig. \ref{fig:7}. This is primarily due to the change in the outer planet's orbital parameters and the reduced resolution of the phase-space map.

\subsubsection{Orbital Stability of Test Particles Beyond the Outer Planet}

We performed test particles simulation to check for a possible stable orbit of a planet in a region exterior to the outer planet. For this, we simulated 21,000 test particles (TPs) scattered between 0 to 8 AU in presence of the inner and outer planets. These planets are set at their best-fit semimajor axis location, but their masses are set to the upper uncertainty limit. The initial eccentricity and inclination of the TPs are set to 10$^{-5}$ and (10$^{-5}$)$^\circ$-60$^\circ$, respectively, and the other orbital parameters ($\omega, \Omega$, and \emph{M}) are randomized between (10$^{-5}$)$^\circ$-360$^\circ$. The TPs do not interact with each other, but they do interact with the known big bodies and evolve due to their gravitational influence.

\begin{figure}[!ht]
\centering
\hspace*{\fill}
\includegraphics[width=1.0\linewidth]{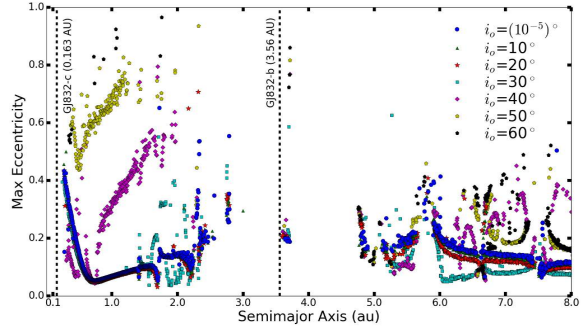}
\hspace*{\fill}%
\caption{Simulation of 21,000 test particles (TPs) randomly placed between 0.163 and 8 AU. The maximum eccentricity of the TPs that survived the total 10 Myr integration are plotted in the $e_{max}$ vs semimajor axis grid. Two vertical dashed lines represent the best-fit semimajor axes of the known planets. Different marker shapes in the legend, from circle to pentagon, represent the TPs that have initial inclination ($i_o$) ranging from (10$^{-5}$)$^\circ$ to 60$^\circ$. Most of the TPs between 0.163 to 3 AU whose $i_o$ distribution is from 50$^\circ$-60$^\circ$, evolve into very high eccentric orbits from initial circular orbits. For orbital inclination between (10$^{-5}$)$^\circ$-40$^\circ$, the maximum eccentricity remains below 0.2 for the majority of the TPs. The majority of the TPs beyond the outer planet maintain stable orbital configuration from 5 to 8 AU for all $i_o$ and their $e_{max}$ remains less than 0.3. It suggests a possibility of having additional planets in the outer region of the outer planet; however, it is not possible to dynamically constrain such bodies and figure out their orbital parameters. Some of the TPs are locked in the mutual co-orbitals with the outer planet, seen along the dashed line at 3.56 AU. They could have been captured by the planet as satellites as well.}
\label{fig:9}
\end{figure}

In Fig. \ref{fig:9} we have plotted the maximum eccentricity versus the semimajor axis of the TPs that survived 10 Myr simulation time. At the beginning of the simulations, the TPs are assigned different initial inclinations ($i_o$), shown in the figure legend. For each $i_o$, the orbits of the TPs evolve in time and, based on their orbital configuration, they either collide with the planets/star, get ejected from the system due to close encounters, or continue to evolve in their orbits. The survival rate of the test particles is high between the known planets (denoted by two vertical dashed line) and exterior to the outer planet, in the region between 5 to 8 AU. Since there are no known bodies beyond the outer planet to constrain the orbital configuration of the test particles, the outward stability region simply continues. Therefore, we chose the cut-off mark at 8 AU.

Most of the surviving TPs between 0.163 to 3 AU whose $i_o$ is between 50$^\circ$-60$^\circ$ evolve into very high eccentric orbits even though their $e_o$ was set close to zero. For higher starting $i_o$ values, the gravitational effects from the existing planets and the star push the TPs into higher eccentric orbits. And, when the $i_o$ is set at relatively less inclined orbits, between (10$^{-5}$)$^\circ$-40$^\circ$, the majority of the test particles' maximum eccentricity remains below 0.2. This complements our analysis in Section 3.2.

The TPs beyond the outer planet maintain stable orbital configuration from 5 to 8 AU for all $i_o$ and their maximum eccentricity remains less than 0.3 for most of the cases. Few test particles are locked in mutual co-orbitals with the outer planet, seen along the dashed line at 3.56 AU. The survival of the test particles in the phase-space between 5 to 8+ AU suggests a potential dynamically stable region for additional bodies. However, any planet residing in this outer region may have minimal to no influence on the observed dynamics of the GJ 832b and the injected middle planet, unless that planet has sufficiently large mass. But, the radial velocity observations of GJ832 b excludes such a scenario. In this paper, we will not explore any further into the dynamical regions beyond the outer planet.


\subsection{Analysis of Synthetic RV Signal}

The orbits of the inner and the middle planets were simulated for 1,500 days, with a high data sampling period (1 per day). The middle planet was first set at 1 AU with an assigned masses of 1 M$_\oplus$, 5 M$_\oplus$ and 10 M$_\oplus$ in a near circular orbits and integrated separately for each mass. Then, using the integrated data, we generated a set of synthetic RV curves based on  the RV signature of Keplerian motion given by equation \ref{eqn:1}, adapted from \citep{sea11}.

The radial velocity equation is given as:

\begin{equation}
V_r = \sqrt{G/(m_1 + m_2)a(1 - e^2)} \cdot m_2sin(i_{sky}) \cdot [cos(\omega + f) + ecos\omega],
\label{eqn:1}
\end{equation}

where, \emph{G} is the universal gravitational constant, \emph{m$_1$} is the stellar mass, \emph{m$_2$} is the planetary mass, \emph{a} is the planet's semimajor axis, \emph{e} is the eccentricity, $i_{sky}$ is the orbital inclination with respect to the sky-plane, $\omega$ is the argument of periapsis, and \emph{f} is the true anomaly.

The caveat of RV technique is that it renders the minimum planetary mass, $msin(i_{sky})$. So, it requires information about the orbital inclination with respect to the sky-plane in order to better constraint its true mass, M$_{true}$ = $M_{min}/sin(i_{sky})$. M$_{true}$ is minimum when $i_{sky}$ = 90$^\circ$. For the RV signal analysis in this section, we have considered the best-fit values as the true mass of the inner and outer planets ( that is, \emph{i} = 90$^\circ$ with respect to the \emph{skyplane}). Also, for the injected middle planets with varying mass, their $msin(i_{sky})$ are assigned 1 M$_\oplus$, 5 M$_\oplus$ and 10 M$_\oplus$, assuming that $i_{sky}$ = 90$^\circ$. Since the planetary mass is the function of its orbital inclination (skyplane), the amplitudes of the RV curves we generated may vary if the \emph{i}$_{sky}$ is less than 90$^\circ$. The data presented here is a special case when the orbital inclination with respect to the sky-plane is maximum for all the planets. Other orbital parameters are the same as the nominal best-fit values given in Table \ref{tab:1}.

\begin{figure}[!ht]
\centering
\hspace*{\fill}%
\includegraphics[width=1.0\linewidth]{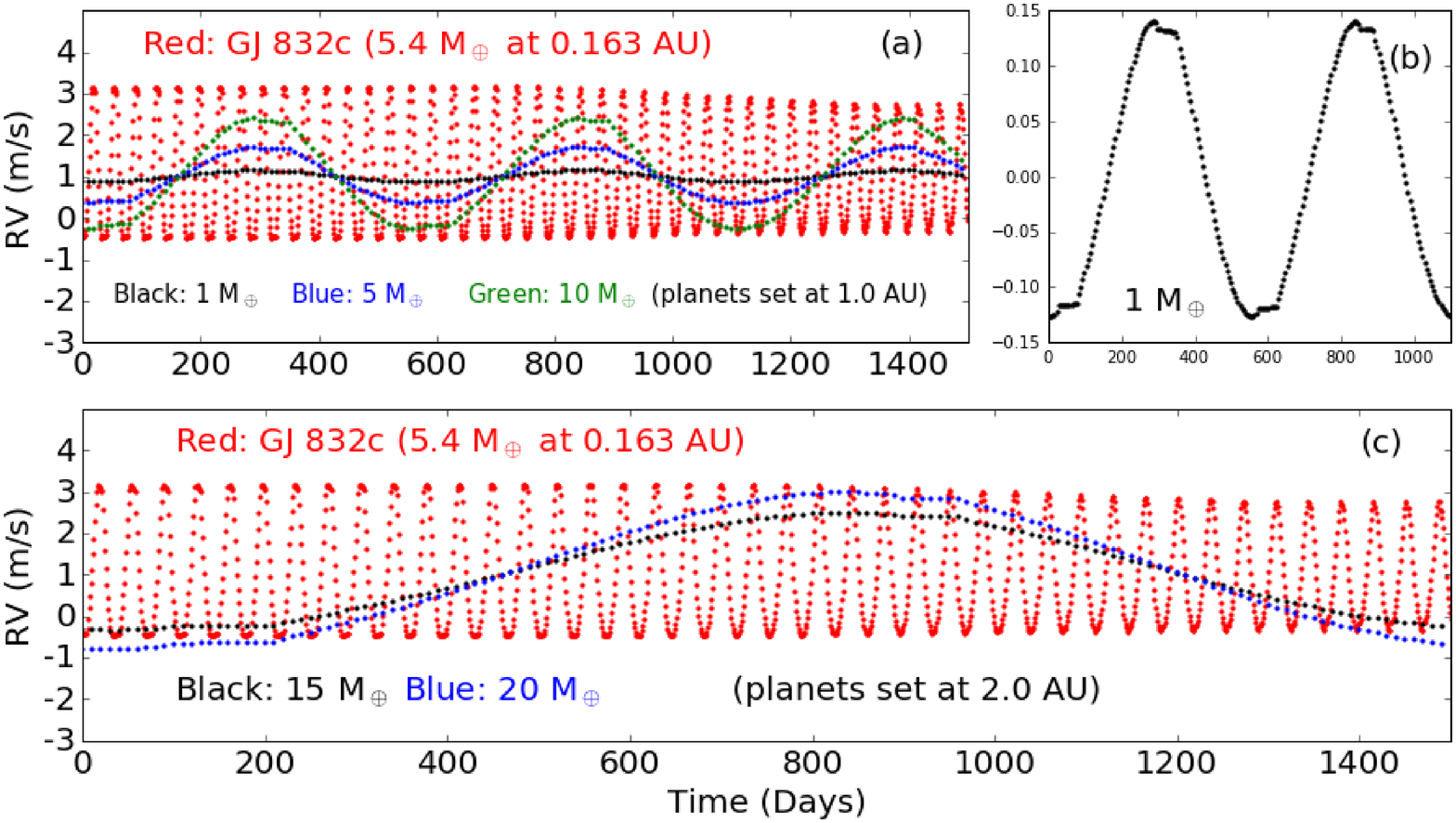}
\hspace*{\fill}%
\caption{(a): Synthetic radial velocity signature of Keplerian motion is shown for the inner planet GJ 832c (red) and the middle planet at 1 AU for 3 different masses for maximum $msin(i_{sky})$. The black, blue and green curves represents the RV signal for 1 M$_\oplus$, 5 M$_\oplus$ and 10 M$_\oplus$, respectively. The maximum amplitude (2.0 m/s) is for the inner planet which is much closer to the star than the injected planet at 1 AU whose amplitude varies by mass: 0.14 m/s , 0.7 m/s  and 1.04 m/s for masses 1 M$_\oplus$, 5 M$_\oplus$ and 10 M$_\oplus$, respectively. The RV indicates that the orbital period for these planets to be about $\sim$550 days. (b): Synthetic RV signal plotted for  1 M$_\oplus$ with smaller y-axis variation. (c): Synthetic RV signal for the planets injected at 2 AU having masses 15 M$_\oplus$ and 20 M$_\oplus$ with an amplitude 1.50 m/s and 2.1 m/s. The RV signal for the 20 M$_\oplus$ planet exceeds the RV signal from the detected inner planet. So the upper limit for the planet's mass, if it is at 2 AU, is $\sim$15 M$_\oplus$ and the orbital period of $\sim$1400 days.}
\label{fig:10}
\end{figure}

The synthesized RV curves for the inner planet (GJ 832c) and the middle planet for three different masses are plotted in Fig. \ref{fig:10}, top panel (a). The maximum amplitude of the RV signal for the inner planet is $\sim$2.0 m/s, similar to the observational value reported by \cite{wit14}. Additional planets with bigger mass, and beyond the orbits of the inner planet, can produce a higher amplitude RV signal, however, the observation has constrained the RV signal for any new planets to be less than 2.0 m/s. For this reason we limited the planetary true-mass at $\le$10 M$_\oplus$ (for maximum \emph{i$_{sky}$}) when placed at 1 AU because the higher mass planet would produce an RV signal greater than 2.0 m/s. However, it is possible that a planet can have larger true-mass and produce the same RV signal for the i$_{sky}$ significantly less than 90$^\circ$.

The injected planets at 1 AU generate the RV curves with varying amplitude as expected. For the masses 1 M$_\oplus$, 5 M$_\oplus$ and 10 M$_\oplus$, the RV signal is 0.14 m/s (black), 0.70 m/s (blue) and 1.04 m/s (green), respectively (Fig. \ref{fig:10}, a). The RV amplitude for 1 M$_\oplus$ is only  $\sim$0.14 m/s, which is much smaller than the current high accuracy RV precision of about 0.97m/s of the HARPS instruments \citep{may03}. A planet interior to 1 AU could be less than 1 M$_\oplus$ but no greater than 10 M$_\oplus$. The planet bigger than 10 M$_\oplus$ (for example 15 M$_\oplus$) had RVs greater than 2.0 m/s, thus we disregarded the results. Any middle planet at 1 AU would have the orbital period of about $\sim$550 days. The highlighted RV signal for 1 M$_\oplus$ is shown in panel (b), with smaller y-axis variation.

The other two planets with masses 15 M$_\oplus$ and 20 M$_\oplus$ injected separately at 2 AU to obtain their synthetic RV reveal that $\sim$15 M$_\oplus$ is the upper mass limit for the middle planet (Fig \ref{fig:10}, bottom panel (c)) because its RV signal is measured less than 2.0 m/s. We chose to set the injected planet at 2 AU marker because it is the farthest stability region observed in Fig. \ref{fig:7}. The RV signals generated for 15 M$_\oplus$ (black) and 20 M$_\oplus$ (blue) planets are 1.50 m/s and 2.10 m/s, respectively. Therefore, we can exclude a planet with mass bigger than 15 M$_\oplus$. The orbital period for a planet with mass $\sim$15 M$_\oplus$ is close to 1400 days. If the location of an injected planet is farther out towards the outer planet, the probable new planet has relatively higher mass than when it is closer towards the inner planet. At the same time, it is possible that the planet can have a mass bigger than 15 M$_\oplus$ and produce the same RV signal for the \emph{i}$_{sky}$ less than 90$^\circ$.

\subsubsection{Observation Prospect of an Additional Planet}

Now, based on the semi-amplitude values, K = ($V_{r,max} - V_{r,min})/2$, and a single measurement precision of $\sigma$ we can estimate the minimum number of observations required to detect an exoplanet (adapted from \cite{pla15}), and is given by:

\begin{equation}
N_{obs} = 2 \Bigg(SNR \cdot \sigma/K \Bigg)^2
\label{eqn:2}
\end{equation}
 
where the SNR is the detection confidence. The calculated $N_{obs}$ based on the equation \ref{eqn:2} for 5$\sigma$ detection are given in Table \ref{tab:2}. The $N_{obs}$ for 1M$_\oplus$, 5M$_\oplus$ and 10M$_\oplus$ planets at 1 AU is 2500, 103 and 47, respectively; and for 15M$_\oplus$ at 2 AU, it is 23. The $N_{obs}$ increases significantly for higher $\sigma$ detection.


\section{Summary}
Our studies of the \emph{lifetime} maps, $e_{max}$ maps, and the time evolution of the orbital elements for GJ 832c establishes the stable orbital configuration for its best-fit orbital solutions. The planet's eccentricity deviation remained within the best-fit uncertainty values during the total simulation time. Based on the $e_{max}$ maps for the phase-spaces in $i_{pl}$, $e_{pl}$ and $a_{pl}$, the relative inclination of the planet is less than 39$^\circ$. Also, the planet remains in stable orbits while maintaining low enough eccentricity deviations during the total integration period of 1 Gyr. The outer planet's orbital elements displayed least deviation from their initial values during the one billion years evolution time; however,due to its gravitational influence, the region starting from 2 AU to 3.56 AU remains dynamically unstable. A region similar to the Solar System's Asteroid belt is likely to exist in the vicinity of 2 AU where the 3:1, 5:2, 7:3, and 2:1 resonances with the outer planet are observed.
 
GJ 832c maintained stable orbit for the $i_{pl}$ $\le$39$^\circ$ and the orbit did not vary significantly even when the middle planet with mass 15 M$_\oplus$ was injected into the system and integrated for 1 Gyr. The middle planet also remained in a stable orbital configuration for the orbital inclination as high as $\sim$30$^\circ$ and for the semimajor axis ranging from 0.75 to 1.25 AU. The $e_{max}$ attained by the middle planet around the 1 AU marker remained close to its initial value even after 1 Gyr orbital evolution. In general, based on the small variations in the initial values of the orbital elements, we could extrapolate our stability analysis and claim that the system is likely to maintain a stable orbital configuration for longer than 1 Gyr timescale.

The injected middle planet could be smaller or bigger than one Earth mass. However, its upper mass limit is constrained by the RV signal of the known inner planet, 2.0 m/s for maximum \emph{i}$_{sky}$. Using this RV signal as a constraint, the synthetic RV of the middle planet is generated from our simulation data for its varying \emph{mass} and $i_{pl}$ for the maximum \emph{i}$_{sky}$. Our results suggest that if the middle planet is located in the vicinity of the 1 AU marker, it has upper mass limit of 10 M$_\oplus$ and generates a RV signal of 1.4 m/s. The 1 M$_\oplus$ planet at the same location has RV signal of 0.14 m/s only, much smaller than the sensitivity of available technology. Nonetheless, the detection using RV method is possible but with a significantly large number of observations: $\sim$2500 for 1 M$_\oplus$ planet. The number of observations can be lowered to 47 if the planet has a mass of 10 M$_\oplus$. The $N_{obs}$ depends on the preferred $\sigma$ detection value as well. Similarly, When the middle planet was fixed to 2 AU, the upper mass limit increased to 15 M$_\oplus$ with synthetic RV signal of 1.5 m/s. Hence, we expect a planet with a mass less than 15 M$_\oplus$ orbiting between the inner and the outer planets. Our RV signal calculation considers only 2 degree variation in \emph{i}$_{sky}$. If the \emph{i}$_{sky}$ varies larger, the mass of the planet will vary according to $msin(i_{sky})$. The orbital period of the planet at 1 AU and 2 AU are $\sim$500 and $\sim$1400 days, respectively. Our aim here is just to provide a general idea of the detection probability to the RV observation scientists. The Earth, for example, exerts 9 cm/s wobble in the Sun. Thus, to detect any Earth-mass planets, the sensitivity of the RV measurements should be down to a few centimeters and require extreme precision radial velocities \citep{fis16}.

The lower stability limit for an Earth-mass planet starts at 0.25 AU and the upper limit of the star's classical habitable zone (HZ) ends at 0.28 AU (from \cite{kop13}). Hence, there is a slim window of about 0.03 AU where an Earth-mass planet (or $\le$15 M$_\oplus$) could be stable as well as remain in the upper limit of the stellar HZ. However, a planet residing in a HZ does not necessarily imply that it can support \emph{life} as we know it. For example, the inner planet, GJ 832c, itself orbits around the lower limit of the stellar HZ but it is not expected to be habitable; see \cite{wit14} for detailed analysis of GJ 832c as a habitable-zone super-Earth.

The upper limit of the planetary mass of GJ 832c is in the range of super-Earths (exoplanets bigger than Earth but smaller than Uranus (15 M$_\oplus$) and Neptune (17 M$_\oplus$)) and has close proximity to the star. These super-Earths can have two formation scenarios: they can form far out and migrate inward to their current location or \emph{in situ} formation which is possible if the planetary disk contains a low turbulence region \citep{mar16}. Also, \cite{izi14} recently found that fast-migrating super-Earths have a modest effect on proto-planetary embryos and planetesimals leaving enough materials to form rocky, Earth-like planets. Hence, if the planet GJ 832c was a fast-migrating super-Earth, that would have left enough space and matter in between the inner and outer planets to have additional bodies.

Long-term orbital stability, orbital dynamics of various phase-spaces and the synthetic RV signal analysis suggest the possible existence of a planet $\le$ 15 M$_\oplus$ between the inner and outer planets in the GJ 832 system. The anticipated RV signal is much lower than the sensitivity of the RV instruments; however, a significantly large number of RV observations and the transit method, provided that the planets are along the line-of-sight, are the viable options to get the observational verifications. Despite being at close proximity (16 ly), the star has aged enough ($\sim$5.6 Gyr) that the residing planets may not radiate enough in order to be detected from imaging technique. A future space telescope, such as Transiting Exoplanet Survey Satellite (2017), TESS (https://tess.gsfc.nasa.gov/) whose mission is to survey G, K, and M type stars including the 1,000 closest red dwarfs, is one of the best options to explore more about this system. In addition, this system is a good candidate for the Jason Webb Space Telescope (2018), JWST (http://www.jwst.nasa.gov/) to perform spectroscopic analysis of the planetary atmosphere.


\textbf{Acknowledgment} We would like to thank the referee for a very comprehensive report on our paper, which allowed us to greatly improve the original version.  We also would like to thank the Office of Graduate Studies at University of Texas at Arlington and their I-Engage Mentoring Program, which initiated this research project to enhance the undergraduate research program at UTA for J.G. Also, Z.E.M. acknowledges the support of this research by the Alexander von Humboldt Foundation. Special thanks to B. Quarles, M. Cuntz and J. Noyola for their discussions, comments and suggestions; and to Mark Sosebee at the high energy computing facility where most of our numerical simulations were carried out.

\begin{table*}[!ht]
\caption{Best-fit orbital parameters of the GJ 832 system obtained from \cite{wit14,bai09}. Mass of the star = 0.45$\pm$0.05 M$_\odot$. The initial values for Inclination and Longitude of Ascending Node of the planets are set by us.}
\centering
\begin{tabular}{|c||c|c|}
\hline
 Parameters & GJ 832b & GJ 832c \\
\hline \hline
\emph{msini} (M$_\oplus$)  & 216 [188, 245] & 5.4 [4.45, 6.35] \\
\hline
Semi-Major Axis (\emph{a}) & 3.56 AU [3.28, 3.84] & 0.163 AU [0.157, 0.169] \\
\hline
Eccentricity (\emph{e}) & 0.08$^{\circ}$ [0.02, 0.1] & 0.18$^{\circ}$ [0.05, 0.31] \\
\hline
Inclination (\emph{i}) & (0-90)$^{\circ}$ & (0-90)$^{\circ}$ \\
\hline
Longitude of the Ascending Node ($\Omega$) & (10$^{-5}$)$^{\circ}$ & (10$^{-5}$)$^{\circ}$ \\
\hline
Argument of the Periapsis ($\omega$) & 246$^{\circ}$ [224, 268] & 10.0$^{\circ}$ [323, 57] \\
\hline
Mean Anomaly ($\mu$) & 307$^{\circ}$ [285, 330] & 165$^{\circ}$ [112, 218]\\
\hline
Period (P) (Days) & 3657 [3553,3761] & 35.68 [35.65,35.71]\\
\hline
\end{tabular}
\label{tab:1}
\end{table*}

\begin{table*}[!ht]
\caption{Radial Velocity semi-amplitude, K = ($V_{r,max} - V_{r,min})/2$, for varying mass and location of an injected planet; and estimated number of observations for different masses and up to 5$\sigma$ detection.}
\centering
\begin{tabular}{|c|c|c|c|}
\hline
 Injected Planet Mass & Dist. from Star (AU) & RV semi-amplitude (m/s) & No. of Observations \\
\hline \hline
1 M$_\oplus$  & 1 & 0.14 & 2500 \\
\hline
5 M$_\oplus$ & 1 & 0.70 & 103 \\
\hline
10 M$_\oplus$ & 1 & 1.40 & 47 \\
\hline
15 M$_\oplus$ & 2 & 1.50 & 23 \\
\hline

\end{tabular}
\label{tab:2}
\end{table*}

\end{document}